\documentclass[]{aastex631}
\usepackage{graphicx} 

\begin{document}

\title{Detecting exoplanet transits with the next generation of X-ray telescopes}

\author[0009-0009-0960-6280]{Raven Cilley}
\affiliation{Department of Astronomy, University of Michigan, Ann Arbor, MI 48109, USA} 

\author[0000-0002-3641-6636]{George W. King}
\affiliation{Department of Astronomy, University of Michigan, Ann Arbor, MI 48109, USA} 

\author[0000-0002-5466-3817]{L\'{i}a Corrales}
\affiliation{Department of Astronomy, University of Michigan, Ann Arbor, MI 48109, USA} 

\begin{abstract}
\label{sec:abstract}
Detecting exoplanet transits at X-ray wavelengths would provide a window into the effects of high energy irradiation on the upper atmospheres of planets. However, stars are relatively dim in the X-ray, making exoplanet transit detections difficult with current X-ray telescopes. To date, only one exoplanet (HD~189733~b) has an X-ray transit detection. In this study, we investigate the capability of future X-ray observatories to detect more exoplanet transits, focusing on both the \textit{NewAthena}-WFI instrument and the proposed Advanced X-ray Imaging Satellite (AXIS), which provide more light-collecting power than current instruments. We examined all the transiting exoplanet systems in the NASA Exoplanet Archive and gathered X-ray flux measurements or estimates for each host star. We then predicted the stellar count rates for both AXIS and \textit{NewAthena} and simulated light curves, using null-hypothesis testing to identify the top 15 transiting planets ranked by potential detection significance. We also evaluate transit detection probabilities when the apparent X-ray radius is enlarged due to atmospheric escape, finding that $\geq 5$ of these planetary systems may be detectable on the $>4\sigma$ level in this scenario. Finally, we note that the assumed host star coronal temperature, which affects the shape of an X-ray transit, can also significantly affect our ability to detect the planet.
\end{abstract}

\section{Introduction}
\label{sec:intro}

The last few decades have yielded the discovery of thousands of exoplanets, over three-quarters of which were detected via the transit method. The majority of these planets are on short period orbits, making them subject to high levels of stellar irradiation that is unparalleled in our own Solar System. In particular, X-ray and EUV emission from a planet-hosting star can heat the upper atmosphere of a planet such that it expands past its radius of gravitational influence, leading to significant atmospheric mass-loss over time \citep{MurrayClay2009, Owen2012, Martinez2019, Fossati2023}. 
Evidence for extended outflows have been observed directly in absorption by Lyman-$\alpha$ \citep{VidalMadjar2003, Ehrenreich2008, Lecavelier2010, Bourrier2013, Ehrenreich2015, Bourrier2018}, weakly ionized metal lines \citep{Fossati2010, Linksy2010, Sing2019, Salz2019}, and metastable He \citep{Spake2018, Allart2018, Nortmann2018, Salz2018, Allart2019, Paragas2021, Fossati2023, Perez2023, Levine2024}. 

The demographics of short period planets to date provide interesting features for study such as the radius gap, a bimodal distribution of planet radii with a sizeable valley in the 1.5-2 Earth radii range \citep{Fulton2017, Fulton2018, VanEylen2018, MacDonald2019}, and a dearth of Neptune-sized planets on very short ($\leq 10$~day) orbits \citep{Mazeh2016}. 
It is hypothesized that atmospheric escape could explain why close-in Neptunes very rarely survive \citep{OwenLai2018} and why sub-Neptunes, which are massive enough to retain a significant atmosphere, are distinctly separate from super-Earths, believed to be rocky cores \citep{Owen2017}. 
Direct observation of escaping atmospheres provides the strongest means by which to test these hypotheses.

Observing transits in the X-ray provides another unique way to observe atmospheric escape. The first and only planetary transit to be detected in the X-ray is that of HD~189733~b \citep[][Wheatley et al., in prep]{Poppenhaeger2013}, one of the earliest discovered planets and a canonical hot Jupiter. The results of \citet{Poppenhaeger2013} imply that the apparent radius of the planet is 50-80\% larger in the X-ray than it is in the optical. This can be attributed to an extended envelope of near-neutral material, for which X-ray absorption via the photoelectric effect is much stronger than typical absorption features in the UV, optical, and infrared -- with the exception of Lyman-$\alpha$. However, the utility of the Lyman-$\alpha$ feature is made difficult by absorption by the intervening interstellar medium. Additionally, interpreting the physical extent and mass-loss rates of planetary outflows from metal line and metastable He absorption is complicated by our lack of knowledge of the exact atmospheric conditions and (variable) radiation field that populates the relevant atomic states \citep{Oklopcic2019, Wang2021a}. X-rays interact with inner-shell electrons, making the photoelectric absorption cross-sections across a broad energy range relatively insensitive to the exact ion population. Thus X-ray transits provide a more agnostic tracer of the physical extent of planetary atmospheres with which to measure atmospheric escape rates, making exoplanet transits a unique subject of interest for future X-ray observatories.

Due to the relative dimness of stars in X-rays, there is no better X-ray transit target than HD~189733~b, which orbits a relatively X-ray bright \citep[$F_{X} \approx 0.5-7\times10^{-13}$ erg\,cm$^{-2}$\,s$^{-1}$;][]{Pillitteri2011, Bourrier2020, Pillitteri2022HD189} K-type star and has a deep $2\%$ level transit \citep{Bouchy2005, Stassun2017}. 
In this study, we investigate the capability of future generations of X-ray telescopes with higher sensitivity, such as AXIS and \textit{NewAthena}, to observe transits of other planets. 
The Advanced X-ray Imaging Satellite (AXIS) is a proposed NASA probe-class mission, designed to be a premier X-ray observatory of the 2030s. The AXIS design features a light-collecting power that makes it 5 times more sensitive to soft X-ray emission from stars than XMM-Newton, and  $\geq 20$ times in the case of Chandra-ACIS \citep{Mushotzky2019, Reynolds2023}. 
The \textit{NewAthena} mission, led by the European Space Agency (ESA), is a large class mission hosting two instruments: the Wide Field Imager (WFI), which provides high sensitivity and a large field-of-view, and X-ray Integral Field Unit (X-IFU) instrument, which provides high spectral resolution across the entire image via micro-calorimeter transition edge sensor technology \citep{Barcons2017}. 
The ability of the X-IFU to resolve spectral changes across a planetary transit are discussed in \citet{Foster2022}. In this work, we focus only on the ability of the WFI Fast Detector to observe a planet transit in the integrated 0.2-2.4~keV soft X-ray band-pass. 
For the sake of comparison to future missions, the integrated effective areas of the current AXIS and \textit{NewAthena}-WFI across this band-pass are 0.8~m$^2$ and 1.4~m$^2$, respectively.

\section{Methods}
\label{sec:methods}
To identify which exoplanet transits may be detectable by AXIS and \textit{NewAthena}, we first collected a large sample of known transiting exoplanets. We drew from the NASA Exoplanet Archive\footnote{\url{https://exoplanetarchive.ipac.caltech.edu/}} (NEA) list of confirmed exoplanets with a detected transit as of May 20th, 2024, using the default data entries for each planet. We excluded any early A-type stars as they lack a convection zone and therefore do not produce coronal/chromospheric emission \citep[e.g.][]{Simon2002,Neff2008,Gunther2022}. We chose to examine all 224 stars with distances less than 100 pc. To avoid missing exceptionally bright systems that are further away, we also considered stars over 100 pc away if they were younger than 1 Gyr. We estimated the X-ray flux of these stars by assuming that they are in the saturated regime, when the ratio of X-ray and bolometric luminosity is $\sim10^{-3}$ \citep{Wright2011, Wright2018}. We then added to our sample any stars which had an estimated X-ray flux greater than $10^{-14}$ erg\,cm$^{-2}$\,s$^{-1}$, of which there were 58. We also added the WASP-180 system into our sample \citep[1.2$\pm1$ Gyr old and 254 pc away, see][]{Temple2019}, as it was listed as an interesting target for future X-ray study by \citet{Foster2022}. This process yielded a list of 283 stars with 416 transiting planets between them. The stars in the sample had stellar types ranging from A9 to M7.5. 

\subsection{X-ray Luminosities and Fluxes}
\label{ssec:xflux}

We first investigated the X-ray emission of each star. We searched for existing measurements of our sample stars from XMM-Newton, Chandra, ROSAT, and eROSITA and examined 22 stellar X-ray studies in the literature which report X-ray observations of at least one star in our sample. We additionally queried the 4XMM-DR12, XMMSSC, CXOGSGSRC, and XMMSLEWFUL catalogs. In total we found X-ray flux measurements for 73 stars in our sample.

For stars with no previous X-ray measurement, we estimated the X-ray luminosity based on empirical relationships between X-ray emission and stellar age or rotation period. The magnetic field of a star, which is strongly connected to the coronal X-ray emission \citep[e.g.][]{Erdelyi2007, Aschwanden2005}, is hypothesized to be powered by varying rotational velocities in the convection zone of the star \citep{Parker1955, Berdyugina2005}. A star's internal differential rotation is dependent on many conditions, including the spectral type, stellar rotation period, and age, as magnetic braking causes a star to spin down over time \citep{Weber1967, Skumanich1972}. 
Thus, X-ray emission is observed to be associated with a star's rotation period and age, a relationship which has been studied for decades \citep[e.g.][]{Rengarajan1984, Noyes1984, Pizzolato2003, Engle2023}. 

Below the age of $\sim$0.1 Gyr (up to $\sim$Gyr for M stars), stellar X-ray emission is in the saturated regime, exhibiting a relatively constant $L_x/L_{bol}$ independent of rotation period \citep{Vilhu1984, Jackson2012}. In the unsaturated regime, the decrease in stellar X-ray emission with age is described by a power law that is different for each spectral type.
Two canonical studies, \citet[][henceforth referred to as J12]{Jackson2012} and \citet[][henceforth referred to as W11 and W18]{Wright2011, Wright2018}, explore the relationships of decreasing X-ray emission with age or rotation period. Both of these studies group stars by color as a proxy for stellar type: W11 and W18 group by V-$K_s$ magnitude, and J12 groups by B-V magnitude. Each color group then has a different power law function that describes the X-ray emission in the unsaturated regime. We utilized these studies to estimate the X-ray properties of the stars in our sample, using B, V, and $K_s$ magnitudes from the SIMBAD database \citep{Wenger2000}.

We prioritized the relations of W18, which uses stellar rotation period as an estimator of X-ray emission, since stellar rotation periods typically have smaller uncertainties than those of stellar ages. 
The relationship in W18 also applies to a larger portion of the stellar types in our sample: W18 includes stellar types ranging from F to M stars $(1\leq(V-K_s)\leq7)$, whereas J12 focuses on F, G, and K stars $(0.29\leq(B-V)<1.41)$. 
For stars without a measured rotation period but with a known age, we estimated the X-ray luminosity using the J12 relation. In the case that a star had no rotation period or age measurement, we chose an age at random from a normal distribution with a mean of 5 Gyr and a standard deviation of 1.5 Gyr, and estimated its X-ray luminosity using the J12 relations. This method assumes those stars to be an average field age while permitting some spread in their values. Nineteen of the stars in our sample have no measured stellar rotation period and were slightly redder ($1.41\leq(B-V)<1.5$) than the B-V value cutoff for the J12 relation, so we grouped them into the closest possible B-V category. 32 stars with no measured rotation period were further outside the B-V cutoff,
so we excluded them from our final sample. 
We excluded an additional 7 stars due to a lack of photometric data reported in SIMBAD. 
To convert X-ray luminosities to unabsorbed flux, we used distances from \textit{Gaia} DR3 \citep{Gaia2016, Gaia2022, Babusiaux2022}. 

This process left us with a final sample of 238 stars, 73 of which have measured X-ray fluxes reported in the literature. Of the 165 stars with estimated fluxes, 67 were based on the 
W18 stellar rotation period relation, 54 on the J12 stellar age relation using a measured age, and 44 on the J12 relation using a randomized age. 
The unabsorbed X-ray fluxes of most stars in our sample were between $10^{-16}$ and $10^{-11}$ erg\,cm$^{-2}$\,s$^{-1}$. 
We report the X-ray luminosities, unabsorbed fluxes, and distances for each host star we examined in this study in a machine-readable tabel in the online Journal; a subset of the data is shown in Table~\ref{tab:sources} as an example.

\begin{deluxetable*}{ccccccc}[hbt!]
\tablecaption{Sample of X-ray Information and Sources table.}
\label{tab:sources}
\tablehead{\colhead{Name} 
    & \colhead{D (pc)} 
    & \colhead{D Ref.\tablenotemark{a}} 
    & \colhead{$L_x$~(erg~s$^{-1}$)\tablenotemark{b}}
    & \colhead{$F_x$~(erg~s$^{-1}$~cm$^{-2}$)\tablenotemark{b}}
    & \colhead{$L_x$ Ref.} & \colhead{$L_x$ Flag\tablenotemark{c}}
    }
\startdata
55 Cnc   & 12.6         & 1 & 3.88E+26       & 2.05E-14      & 2011A\&A...532A...6S  & 0\\
AU Mic   & 9.71          & 1 & 4.80E+29       & 4.26E-11       & 1999A\&AS..135..319H & 0\\
DS Tuc A & 44.2         & 1 & 2.14E+30       & 9.17E-12       & 2021A\&A...650A..66B & 0\\
G 9-40   & 27.8         & 1 & 3.62E+27       & 3.91E-14       & 2020AJ....159..100S  & 1\\
GJ 1132  & 12.6         & 1 & 1.03E+26       & 5.43E-15       & 2020A\&A...641A.136W & 0\\
\enddata
\tablenotetext{a}{Reference 1 indicates Gaia DR3 (2016A\&A...595A...1G, 2023A\&A...674A...1G)}
\tablenotetext{b}{X-ray luminosities and unabsorbed fluxes are reported for the 0.2-2.4 keV band.}
\tablenotetext{c}{Indicates which method was used to obtain the X-ray luminosity. 0 corresponds to a measured X-ray flux, and 1 corresponds to a luminosity estimated using the stellar rotation period.}
\end{deluxetable*}

\subsection{Light Curve Simulations}
\label{ssec:lcurve}

We used XSPEC version 12.11.1d \citep{Arnaud1996} to estimate an AXIS and \textit{NewAthena} count rate for each star. For stars with flux measurements in the literature, we used the same model as reported by those papers to find the count rate. For stars with no model given or no known measurement, we adopted a one-temperature APEC model \citep{Smith2001} for the coronal plasma. 
We chose the coronal plasma temperatures based on the age of the star, because the decrease in X-ray emission with age also correlates with a decrease in coronal temperature \citep[e.g.][]{Schmitt1990}. 
We used a temperature of 0.5 keV for stars younger than 1 Gyr and 0.25 keV for the older stars. We also used the \texttt{tbabs} model \citep{Wilms2000} to account for interstellar X-ray absorption. 
Following \citet{Redfield2000}, we estimated the interstellar hydrogen column using the formula ${\rm NH} \approx d \times 0.1~{\rm cm}^{-3}$, where $d$ is the distance to the star system. 
We ran the \textit{fakeit} command to simulate each spectrum, with an exposure of one million seconds to ensure each target had a statistically significant number of counts, and used this to estimate the count rates for each star in AXIS and \textit{NewAthena}.

X-ray emission from F-type stars and later arises from the optically thin corona, causing the edges of the stellar disk to appear brighter in X-rays than the center. This required us to use a unique limb brightening (rather than limb darkening) function. 
To incorporate limb brightening into our model light curves, we created a custom version of the \texttt{batman} python package \citep{Kreidberg2015}. Our model assumes radial symmetry of the corona about the star and an exponentially decaying coronal intensity dependent on the pressure scale height of the corona (King et al., submitted). We estimated the pressure scale heights for each target star under the assumption of an isothermal plasma
with the equation
\begin{equation}
    \label{eq:H}
    H = \frac{2k_BT}{\mu m_Hg}
\end{equation}
where g is the surface gravity of the star, $k_B$ is the Boltzmann constant, $\mu$ is the mean particle weight of the corona (we used $\mu=1.27$), and $m_H$ is the mass of the hydrogen atom. 

We modeled each transit using planetary parameters from the NEA, assuming that the apparent radius of the planet in X-rays $(\frac{R_p}{R_*})$ is equal to that measured in the optical. We initiated each \textit{batman} transit model at a high time resolution, which we then integrated to obtain 500 second bins.
Each transit model spans 30 ks (about 8.3 hours), yielding 60 model data points in each light curve. 
We simulated each instrumental light curve by multiplying the \texttt{batman} transit model by the out-of-transit counts per bin expected for the system in question, then drawing data points for each bin from a Poissonian distribution.
Finally, we renormalized the synthetic light curve by the out-of-transit number of counts per bin 
to allow for visual comparison among the light curves of different transiting exoplanets. 

\begin{figure}[htb!]
    \centering
    \includegraphics[scale=0.3]{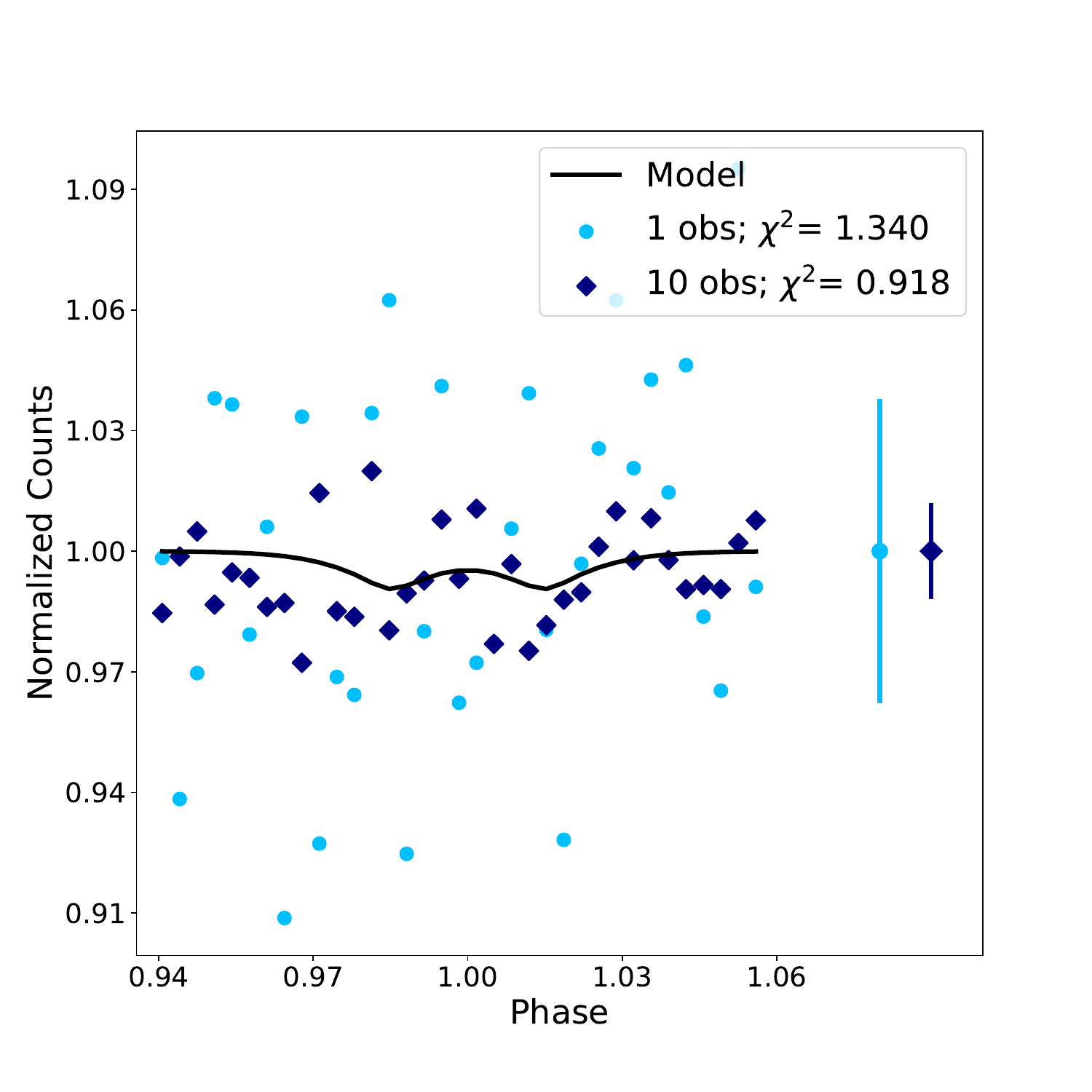}
    \caption{Simulated transit light curve data sets, for one transit (blue circles) and 10 averaged transit observations (blue diamonds) of WASP-93 b. The 10-averaged data set has less statistical scatter and is much closer to the transit model (black line) than that of the single set of simulated data. The two points to the right of the light curve show the standard deviation of each data set.}
    \label{fig:Stacking}
\end{figure}

We also experimented with averaging transit observations, which reduces scatter and, in practice, will be necessary to reduce the effects of stellar variability \citep{Llama2015, Llama2016}. 
We simulated a multi-transit observation by generating 10 light curves and averaging the counts in each time bin. Figure~\ref{fig:Stacking} provides a comparison of simulated data for 10 averaged observations versus only one observation of the WASP-93 b transit. 
For both AXIS and \textit{NewAthena}, we found that 
with only one observed light curve, none of the transits could be detected with $\geq1-1.5 \sigma$ significance following the procedures outlined in Section~\ref{sec:Detectability}. However, by combining 10  light curves ($\approx 300$~ks for most short period planets) a robust transit detection could be obtained for some targets. All procedures and results described below utilize the simulated light curves stacked from 10 hypothetical transit observations.

\subsection{Determining detection likelihood}
\label{sec:Detectability}

For each planetary system, we determined whether or not an X-ray transit is detectable, and to what statistical significance, by employing the null hypothesis testing procedure described in \citet{XrayHandbook}. First, we 
computed the likelihood value ($L_t$, for which we used the $\chi^2$ function) for the input transit model to fit the simulated transit dataset. 
Then we fit the light curve with a constant (no-transit) model and computed the likelihood ($L_0$) for the null-hypothesis. 
Taking the ratio of the likelihood of these two models provides a rough estimate for which model is more likely to be adopted as correct. For the simulated transit dataset, we label the likelihood ratio as $(L_t/L_0)_{d}$. 
We then produced one million simulated datasets assuming no transit (a constant count rate) and computed the likelihood ratios for the null-hypothesis models, $(L_t/L_0)_n$, following the same procedures.

Comparing the $(L_t/L_0)_{d}$ value with the distribution of $(L_t/L_0)_{n}$ provides a probability of a false-positive detection (p-value), i.e., the probability that the transit model would appropriately fit a light curve that has no transit in it. We computed the p-value by calculating the percentile of $(L_t/L_0)_{n}$ on which the $(L_t/L_0)_{d}$ value fell. This analysis mimics what an observer might do to verify a transit detection. 
We performed this analysis for each of the potential exoplanet targets, ranking each transit system based on its p-value. 
Since the $(L_t/L_0)_{n}$ distributions were comprised of one million values, 
we did not have the appropriate number of samples to determine very small p-values $\lesssim 10^{-6}$. 
We chose a p-value cut-off of $10^{-5}$, corresponding to a $4.4\sigma$ detection, so that any calculated p-value that fell below this value 
was classified as a probable $\geq 4.4 \sigma$ detection. 

Random noise contributed significantly to the apparent detectability of the simulated transits. 
To address this, 
we repeated the p-value calculation five times for each of the potential 238 transit targets. We used the median p-value of these samples to rank the entire list, 
from which we selected the top 25 viable detections. Of those top 25, we repeated the p-value calculation 10 more times and took the median value, 
from which we selected the top 15. 
For our final listing of the top 15 X-ray transit targets, we repeated the process for 100 simulated transit datasets, reporting the median p-value in rank order (Tables~\ref{tab:TopTargetsAXIS} and \ref{tab:ATHENA}).

\begin{figure*}[htb!]
   \centering
    \includegraphics[scale=0.5]{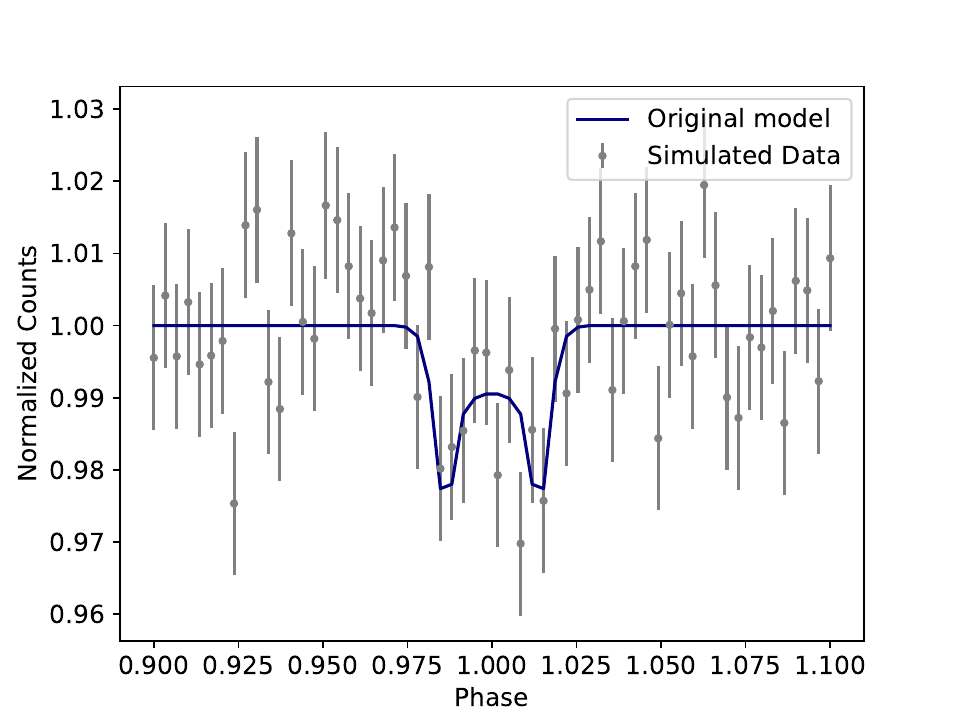}
    \includegraphics[scale=0.6]{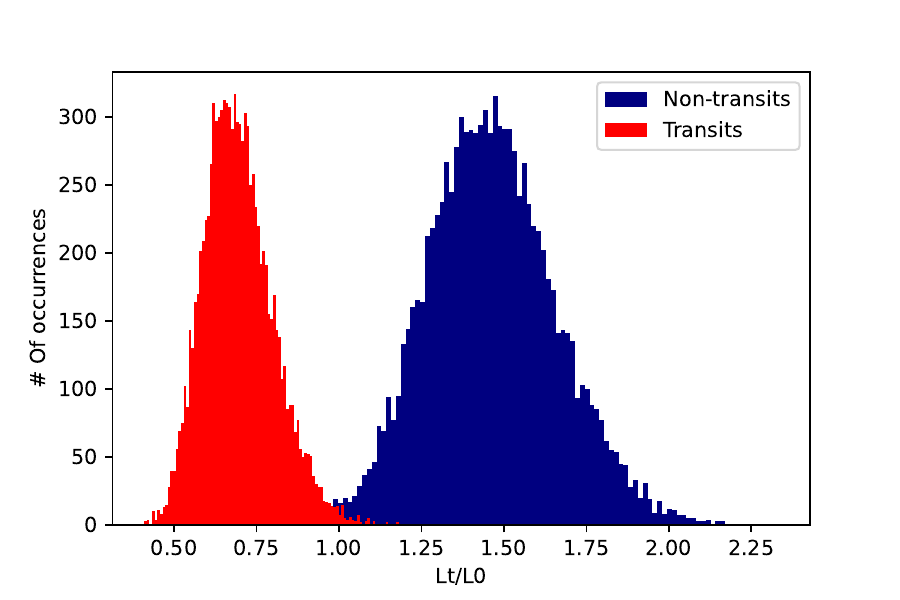}
    \caption{HD 189733 b example simulated transit light curve and null-hypothesis analysis. Left: HD 189733 b transit model (solid blue curve) and simulated data for 10 averaged AXIS observations (gray). 
    Right: The $(L_t/L_0)_{n}$ distribution generated for $10^4$ sets of data with no  transit, and a distribution of $(L_t/L_0)_{d}$ values for $10^4$ sets of simulated AXIS HD~189733~b transit data. The median p-value is $\leq10^{-5}$, indicating that HD 189733 b could be detected at $\geq4.4\sigma$ significance with AXIS. Our method for analyzing transit light curves is described in section \ref{sec:Detectability}.}
    \label{fig:HD189LightCurves} 
\end{figure*}

Figure~\ref{fig:HD189LightCurves} shows an example of the difference between the distribution of $(L_t/L0)_d$ and $(L_t/L_0)_n$ in our null hypothesis analysis for HD~189733~b. For a given median p-value (and associated level of significance $n$, in units of $\sigma$), our experiment suggests that there is a 50\% chance that the transit observing campaign will yield a transit detection to $> n\sigma$. In many instances, there is significant overlap between the $(L_t/L_0)_d$ and $(L_t/L_0)_n$ distributions, so that some measured p-values imply $\leq 1\sigma$ significance, which we label as a non-detections. The total number of trials resulting in non-detections are reported in both Tables~\ref{tab:TopTargetsAXIS} and Tables~\ref{tab:ATHENA}, providing an estimate of the percentage risk that a transit observing campaign will fail to detect the planet.

\subsection{Re-analysis of HIP~65~A X-ray Luminosity}
\label{ssec:HIP65}
HIP 65 A b is an ultra-short period Jupiter which orbits a K4 dwarf star \citet{Nielsen2020}. It was initially the top ranked candidate target in our sample for both observatories, due to the high stellar X-ray luminosity measurement reported by the 4XMM-DR12 XMM-Newton Serendipitous Source Catalog \citep{Webb2020}. This was the only system we categorized as potentially observable with a stellar X-ray luminosity measurement from a catalog with a `one-size-fits-all' model approach. To check the validity of this measurement, we re-analyzed the XMM-Newton data directly (Obs ID: 0862641401; PI: Mantz). 

We reduced the XMM-Newton EPIC-pn data using the Science Analysis System (SAS; \citet{SAS}) tools provided by the XMM-Newton science team, following the standard procedure outlined in the SAS Threads\footnote{\url{https://www.cosmos.esa.int/web/xmm-newton/sas-threads}}. High-energy background radiation driven by solar soft proton flares \citep{Walsh2014} was prominent in the observation, so we filtered out about three-quarters of the observation for our spectral analysis. We then used XSPEC to fit the observed spectrum, utilizing a 1-temperature APEC model and a \texttt{tbabs} model (parameters given in Table~\ref{tab:HIP65Params}). We used solar abundances from \citet{Aschwanden2005} and the C-statistic \citep{Cash1979} for fitting the spectral model. We also inspected the light curve visually and found no obvious flaring. 

The final unabsorbed flux we obtained was $3.97^{+0.81}_{-0.72}\times10^{-14}$ erg\,cm$^{-2}$\,s$^{-1}$ (0.2-2.4 keV), which was approximately one third of our original value. 
We re-ran the transit analysis for HIP~65~A~b using this flux value, which is reported in the final results. 
Even with the lower flux value, HIP~65~A~b still maintained its position in the top 15 candidate targets, moving from first to second place. 
This is due to the extreme transit depth of the system, which is around 8\% in the optical \citep{Nielsen2020}.

\begin{deluxetable*}{ccccc}[!htb]
\tablecaption{XSPEC model fit parameters for HIP~65~A.}
\label{tab:HIP65Params}
\tablehead{
    \colhead{NH} & 
    \colhead{kT} & 
    \colhead{EM} & 
    \colhead{$L_X$} & 
    \colhead{$F_X$} \\
    ($10^{20}$ cm$^{-2}$) & 
    (keV) & 
    ($10^{51}$ cm$^{-3}$) & 
    ($10^{28}$ erg\,s$^{-1}$) & 
    ($10^{-14}$ erg\,cm$^{-2}$\,s$^{-1}$)
    }
\startdata
    $0.19$ (fixed) & 
    $0.28^{+0.05}_{-0.03}$ & 
    $1.04^{+0.22}_{-0.20}$  & 
    $1.82^{+0.37}_{-0.33}$ & 
    $3.97^{+0.81}_{-0.72}$      \\    
\enddata
\end{deluxetable*}

\section{Results}
\label{sec:results}

\begin{deluxetable}{lcccc}[htb!]
\tablecaption{Top AXIS transit candidates}
\label{tab:TopTargetsAXIS}
\tablehead{\colhead{Planet} & \colhead{Med p-value} & \colhead{Sig ($\sigma$)} & \colhead{N $\leq1\sigma$} & \colhead{ct~s$^{-1}$}}
\startdata
HD 189733 b & $<10^{-5}$ & $\geq4.4$ & 0 & 1.97 \\
HIP 65 A b  & 2.5$\times$10$^{-3}$ & 3.0  & 2 & 0.16     \\
WASP-93 b$^r$  & 0.01 & 2.4  & 1 & 1.40     \\
AU Mic b    & 0.04 & 2.0 & 11 & 166     \\
WASP-140 b & 0.16 & $<1.5$ & 35 & 0.12     \\
WASP-135 b$^a$  & 0.13 & 1.5 & 26 & 0.16    \\
TOI-620 b$^r$ & 0.23 & $<1.5$ & 37 & 2.44    \\
HIP 67522 b & 0.32 & $<1.5$ & 50 & 6.79    \\
WASP-80 b   & 0.25 & $<1.5$  & 40 & 0.12     \\
DS Tuc A b & 0.45 & $<1.5$  & 57 & 31.3     \\
WASP-180 A b & 0.29 & $<1.5$ & 47 & 0.33 \\
WASP-77 A b & 0.52 & $<1.5$  & 61 & 0.04    \\
WASP-43 b  & 0.40 & $<1.5$ & 59 & 0.03       \\
WASP-52 b$^a$  & 0.39 & $<1.5$ & 60 & 0.06     \\
WASP-145 A b$^a$ & 0.60 & $<1.5$ & 66 & 0.04    \\
\hline
\multicolumn{3}{l}{$^r$Fluxes estimated from stellar rotation period} \\
\multicolumn{3}{l}{$^a$Fluxes estimated from stellar age}
\enddata
\tablecomments{
The median p-value, and corresponding level of significance in units of $\sigma$, is derived from 100 simulated transit datasets, which  each consist of 10 stacked light curves. 
We count the number of simulations yielding a p-value corresponding to $\leq 1 \sigma$ significance as a non-detection.
}
\end{deluxetable}

Table~\ref{tab:TopTargetsAXIS} lists the 15 transiting exoplanets most likely to be observable by AXIS. Our results demonstrate that AXIS is likely capable of achieving a $4.4\sigma$ or greater detection of the HD~189733~b transit. The next two targets on the list, HIP~65~A~b and WASP-93~b, with median p-values around 0.003 and 0.01 respectively, are also likely to be marginally detectable: HIP 65 A b to $3\sigma$ significance, and WASP-93 b to $\sim2.5\sigma$ significance. However, in both cases, there is a risk of no transit detection: 2$\%$ of the trials for HIP 65 A b and 1$\%$ of the trials for WASP-93 b. 
AU~Mic~b was the fourth best transiting target primarily due to the high X-ray flux of its host star, though the small $\frac{R_p}{R_*}$ \citep[about 0.05 in optical,][]{Gilbert2022} made the transit more difficult to detect than other top targets. 
The median p-value suggested that a 2$\sigma$ detection may be possible. However, 11\% of the AXIS trials for AU~Mic~b yielded a $\leq$1$\sigma$ detection.
We also note that AU~Mic has the highest estimated AXIS photon count rate in our sample (166 counts/second), which is likely to cause significant non-linear detector effects, such as pile-up, without proper mitigation. 
The remaining transits are not likely to be observable with AXIS at a significance greater than 1.5$\sigma$, as none of their median p-values were less than 0.1 and detections of less than 1$\sigma$ significance were frequent.

The \textit{NewAthena} observatory would provide slight increases in signal-to-noise for transit light curves, as compared to AXIS. 
Table~\ref{tab:ATHENA} lists the median p-value obtained from 100 simulated trials, 
as described for the AXIS trials. \textit{NewAthena} will also likely be able to detect HD~189733~b to at least 4.4$\sigma$ significance. The likelihood for \textit{NewAthena} to be able to detect HIP~65~A~b, WASP-93~b, and AU~Mic~b were slightly higher than those of AXIS.
The median p-values for WASP-135 b and WASP-140 b indicated that $\sim$2$\sigma$ detections may be possible with \textit{NewAthena}, but $\sim$20\% of the trials yielded a null-result for both planets. The remaining targets, while still unlikely to be detected with 10 transit visits, had lower p-values than the corresponding trials with AXIS. 

These conclusions hold when the apparent X-ray radius of each planet is equal to that of the optical. In the next section, we explore the possibility that the apparent X-ray radius of some of the planets is larger due to atmospheric escape. %
We also limit ourselves to examining transit detections with \textit{NewAthena} only, since the soft energy sensitivity is similar to but slightly better than that of AXIS.

\begin{deluxetable}{lcccc}[htb!]
\tablecaption{Top \textit{NewAthena} transit candidates}
\label{tab:ATHENA}
\tablehead{\colhead{Planet} & \colhead{Med p-value} & \colhead{Sig ($\sigma$)} & \colhead{N $\leq 1\sigma$} & \colhead{ct~s$^{-1}$}}
\startdata
HD 189733 b & $<10^{-5}$ & 4.4 & 0  & 2.15 \\
HIP 65 A b  & 6.9$\times$10$^{-4}$ & 3.4 & 0  & 0.19 \\
WASP-93 b$^r$ & 5.9$\times$10$^{-3}$  & 2.8 & 2  & 1.97  \\
AU Mic b    & 0.03 & 2.2 & 8 & 224 \\
WASP-140 b  & 0.08 & 1.8 & 18 & 0.16 \\
WASP-135 b$^a$  & 0.07 & 1.8 & 21 & 0.22 \\
TOI-620 b$^r$ & 0.22 & $<1.5$ & 41 & 2.69 \\
HIP 67522 b & 0.17 & $<1.5$ & 34 & 9.33 \\
WASP-80 b   & 0.21 & $<1.5$ & 40 & 0.12 \\
DS Tuc A b  & 0.28 & $<1.5$ & 43 & 45.8 \\
WASP-180 A b & 0.30 & $<1.5$ & 48 & 0.38 \\
WASP-77 A b & 0.40 & $<1.5$ & 60 & 0.04 \\
WASP-43 b   & 0.28 & $<1.5$ & 49  & 0.03 \\
WASP-52 b$^a$   & 0.31 & $<1.5$ & 49 & 0.09 \\
WASP-145 A b $^a$ & 0.58 & $<1.5$ & 72 & 0.04 \\
\hline
\multicolumn{3}{l}{$^r$Fluxes estimated from stellar rotation period} \\
\multicolumn{3}{l}{$^a$Fluxes estimated from stellar age} \\
\enddata
\tablecomments{
The median p-value, and corresponding level of significance in units of $\sigma$, is derived from 100 simulated transit datasets, which  each consist of 10 stacked light curves. 
We count the number of simulations yielding a p-value corresponding to $\leq 1 \sigma$ significance as a non-detection.}
\end{deluxetable}
\subsection{Extended X-ray Absorbing Atmospheres}
\label{ssec:extendedatmospheres}

Atmospheric photoevaporation can cause the radius of an exoplanet to appear substantially larger in X-rays than in the optical. X-ray observations of HD~189733~b \citep{Poppenhaeger2013} indicate that even moderate amounts of photoevaporation induced mass-loss ($\sim 10^{11}$~g~s$^{-1}$) can lead to X-ray transits that are significantly deeper, $\sim 2-3$ times that of the optical. 
Thus, many X-ray transits may appear deeper and wider in real observations than our transit simulations above, which assumed an $\frac{R_p}{R_*}$ equal to that inferred from optical transits.

To investigate the detectability of our top candidate targets with extended X-ray absorbing atmospheres, we re-ran our \textit{NewAthena} transit simulations for the top 15 systems with planet radii that were 50\%, 100\%, and 150\% larger than optical. We chose to examine some of the largest radii as a limiting case, although they are technically not outside the range of feasibility. As an example, \citet{Ehrenreich2015} found that GJ~436~b, a sub-Neptune planet with a moderate mass-loss rate (similar to HD~189733~b), had a Lyman-$\alpha$ transit depth of over 50\%, implying an effective radius $10$ times the optical. This surprise demonstrates that there is no well-known upper limit to transit depths at extremely short wavelengths.  Following the same methods as described above, we implemented the null hypothesis testing procedure with 100 simulated transit datasets of 10 averaged observations each. Table~\ref{tab:radchange} gives the results. Assuming they have enough atmospheric evaporation to yield a larger effective radius in the X-ray, nearly all of the top 15 candidate X-ray transit targets emerge as great candidates for \textit{NewAthena} observation. 

A radius increase of 50\% led to significantly more transits being detectable at a $\sim 2-3\sigma$ significance: WASP-135 b, HIP 67522 b, WASP-80 b, WASP-180 b, WASP-52 b, DS Tuc A b, and TOI-620 b. The results for WASP-140 b indicated a $\sim$3.5$\sigma$ detection may be possible with \textit{NewAthena}, and HD 189733 b, HIP 65 A b, WASP-93 b, and AU Mic b all had median p-values which indicated a $\geq$4.4$\sigma$ detection is likely to be possible. There were also a few transits which were on the edge of detectable with \textit{NewAthena} assuming a radius 50$\%$ larger than optical: WASP-77 A b and WASP-43 b had median p values which indicated a 1.5-2$\sigma$ detection may be possible.
With X-ray radii twice that of the optical, \textit{NewAthena} is likely to be able to detect 13 of the target candidates to $\geq3\sigma$ significance. For radii that are 2.5 times larger than the optical, all but one of the 15 transit targets examined are detectable to $\geq3\sigma$. Of these, 13 would be detectable to $\geq4.4\sigma$ significance.

\begin{deluxetable*}{lccccccc}[htb!]
\tablecaption{\textit{NewAthena} detection likelihood with various radius multipliers}
\label{tab:radchange}
\tablehead{\colhead{Planet} & \colhead{$(R_p/R_*)_{\rm opt}$} & \colhead{$1.5\times R_{p}$} & \colhead{Sig $(\sigma)$} & \colhead{$2\times R_p$}& \colhead{Sig $(\sigma)$} & \colhead{$2.5 \times R_p$} & \colhead{Sig $(\sigma)$} }
\startdata
HD 189733 b  & 0.16 & $<10^{-5}$ & $\geq$4.4 & $<10^{-5}$ & $\geq$4.4 & $<10^{-5}$ & $\geq$4.4 \\
HIP 65 A b   & 0.29 & $<10^{-5}$ & $\geq$4.4 & $<10^{-5}$ & $\geq$4.4 & $<10^{-5}$ & $\geq$4.4  \\
WASP-93 b$^r$    & 0.11 & $<10^{-5}$ & $\geq$4.4 & $<10^{-5}$ & $\geq$4.4 & $<10^{-5}$ & $\geq$4.4 \\
AU Mic b     & 0.05 & $<10^{-5}$ & $\geq$4.4 & $<10^{-5}$  & $\geq$4.4 & $<10^{-5}$ & $\geq$4.4 \\
WASP-140 b   & 0.14$^t$ & 6.3$\times10^{-4}$ & 3.4 & $<10^{-5}$ & $\geq$4.4 & $<10^{-5}$ & $\geq$4.4 \\
WASP-135 b$^a$   & 0.14 & 2.7$\times10^{-3}$ & 3.0 & $<10^{-5}$ & $\geq$4.4 & $<10^{-5}$ & $\geq$4.4\\
TOI-620 b$^r$    & 0.06 & 0.03 & 2.1 & 1.6$\times10^{-4}$ & 3.8  & $<10^{-5}$ & $\geq$4.4 \\
HIP 67522 b  & 0.07 & 1.9$\times10^{-3}$ & 3.1 &  $<10^{-5}$  & $\geq$4.4 & $<10^{-5}$ & $\geq$4.4\\
WASP-80 b    & 0.17 & 0.01 & 2.5 & 2.7$\times10^{-5}$ & 4.2 & $<10^{-5}$ & $\geq$4.4 \\
DS Tuc A b   & 0.05 & 0.02 & 2.3 & 1.2$\times10^{-4}$ & 3.9 & $<10^{-5}$ & $\geq$4.4 \\
WASP-180 A b & 0.11 & 0.02 & 2.4 & 8.7$\times10^{-5}$ & 3.9 & $<10^{-5}$ & $\geq$4.4 \\
WASP-77 A b  & 0.13 &  0.07 & 1.8 & 0.02 & 2.2 & 1.2$\times10^{-4}$ & 3.8\\
WASP-43 b    & 0.16 & 0.08 & 1.8 & 2.8$\times10^{-3}$ & 3.0 & $<10^{-5}$ & $\geq$4.4  \\
WASP-52 b$^a$    & 0.17 & 0.04 & 2.0 & 4.5$\times10^{-5}$ & 4.1 & $<10^{-5}$ & $\geq$4.4\\
WASP-145 A b$^a$ & 0.11$^t$ & 0.23 & $<1.5$ & 0.04 & 2.0 & 3.5$\times10^{-3}$  & 2.9\\
\hline
\multicolumn{8}{l}{$^r$Fluxes estimated from stellar rotation period} \\
\multicolumn{8}{l}{$^a$Fluxes estimated from stellar age}  \\
\multicolumn{8}{l}{$^t$Reported optical radius of the planet from the NASA Exoplanet Archive} 
\enddata
\tablecomments{Each median p-value is derived from 100 randomized data sets, which each consist of 10 stacked light curves. }
\end{deluxetable*}

\begin{figure}[htb!]
   \centering
    \includegraphics[scale=0.5]{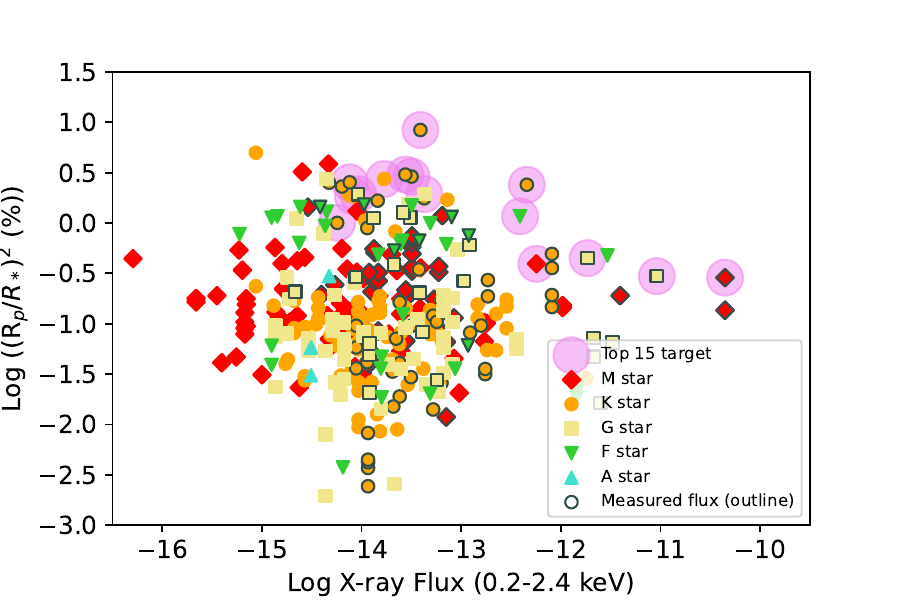}
    \includegraphics[scale=0.5]{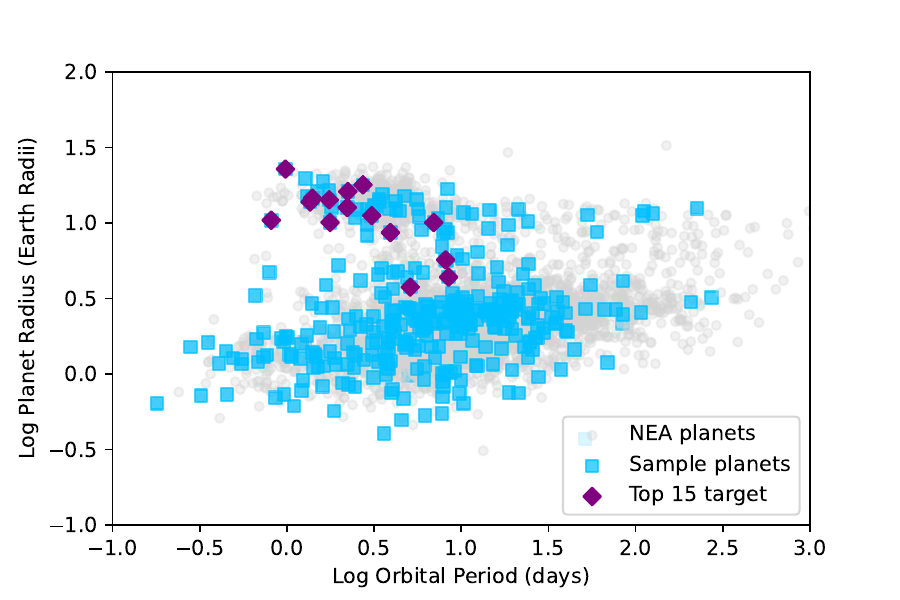}
    \caption{Stellar and planetary properties of our sample. The left plot shows log X-ray flux at Earth (0.2-2.4 keV) vs log $\frac{R_p}{R_*}^2$ ($\%$). Each planetary system in the sample is plotted, with the host star's spectral type determining the color and shape. Points with dark outlines have measured X-ray fluxes, and points surrounded with purple circles are in the list of the top 15 candidate targets for AXIS and \textit{NewAthena}. The method for determining each X-ray flux is given in section \ref{ssec:xflux}. For fluxes given in literature in a band other than 0.2-2.4 keV, we used the XSPEC models described in section \ref{ssec:lcurve} to convert the flux to 0.2-2.4 keV. The right plot shows the orbital period (days) and radii (Earth radii) of the planets in the NEA, with planets in our sample highlighted in blue squares. The top 15 candidate targets are shown with darker blue diamonds. The method for determining the top target candidates is described in Section \ref{ssec:lcurve}.}
    \label{fig:TranDepScatter} 
\end{figure}

\subsection{Variation of Coronal Model and Transit Depth/Shape}
\label{ssec:HVariation}

The stellar coronal scale height, which affects the shape and depth of the X-ray transit (King et al., submitted), could also affect our ability to detect the transit. To investigate the impact of our chosen coronal models, we reran our simulations for the top 4 targets using a set range of scale height values.
In table \ref{tab:ScaleHeight} we show the maximum depth of the X-ray transit model and \textit{NewAthena} p-values of the top 4 transit targets 
using coronal scale heights varying from 0.05 $R_*$ to 0.45 $R_*$. 
In general, increasing the coronal scale height produces a shallower transit, leading to higher p-values and lower detectability. 
Grazing transits, such as HIP~65~A~b, which has an impact parameter of about 1.17 \citep{Nielsen2020}, follow different trends depending on their specific inclinations and other system parameters. For HD~189733~b, the p-value changed by several orders of magnitude depending on the chosen coronal scale height value. This demonstrates that 
a detailed understanding of stellar coronal temperatures and structure is imperative for modeling and interpreting X-ray transits.

\begin{deluxetable*}{lcccccccccc}[htb!]
\centering
\caption{\textit{NewAthena} X-ray transit depths and detection likelihoods from varying coronal pressure scale height values}
\label{tab:ScaleHeight}
\tablehead{\colhead{}  & \multicolumn{3}{c}{H = 0.05$R_*$} & \multicolumn{3}{c}{H = 0.15$R_*$} & \multicolumn{3}{c}{H = 0.45$R_*$}}
\startdata
Planet  & $^{\dagger}\delta_X$ (\%) & Med p & Sig $(\sigma)$   & $^{\dagger}\delta_X$ (\%) & Med p & Sig $(\sigma)$ & $^{\dagger}\delta_X$ (\%) & Med p  & Sig $(\sigma)$\\ \hline
HD 189733 b & 2.66       & $<10^{-5}$ & $\geq$4.4   & 2.03       & $<10^{-5}$ & $\geq4.4$        & 0.97       &  3.8$\times10^{-3}$ &  2.9      \\
HIP 65 A b  & 4.16       & 2.5$\times10^{-3}$  &  3.0  & 4.80       & 2.0$\times10^{-5}$ &  4.3      & 2.96        & 6.5$\times10^{-4}$ &  3.4      \\
WASP-93 b   & 1.66       & $<10^{-5}$  & $\geq$4.4   &  1.17      & 5.8$\times10^{-4}$ & 3.4  & 0.52       & 0.03 &2.1 \\
AU Mic b    & 0.14       & 2.0$\times10^{-3}$ & 3.1   & 0.15       & 2.6$\times10^{-3}$  &  3.0 & 0.09       & 0.07  & 1.8 \\
\hline
\multicolumn{10}{l}{$^{\dagger}$The value of maximum depth in the X-ray transit model}
\enddata
\tablecomments{Each median p-value is derived from 100 randomized data sets, with 10 averaged transit observations in each set. }
\end{deluxetable*}
\subsection{Effects of Stellar Flares on Transit Detection}
\label{ssec:Flares}
Stellar variability presents one of the largest barriers for transit detection in the X-ray. Many planets in our sample orbit active stars that are expected to flare often in the X-ray. Typical stellar X-ray flares are seen as a sudden (factor of several) increase in the X-ray light curve, with a decay time on the order of hours. An observer will discard these flare events from the dataset. To examine how such flares may impact the detectability of transits, we masked sections in simulated \textit{NewAthena} transit light curves of HD~189733~b and AU~Mic~b, two systems in our sample with stars known to be X-ray active \citep{Bourrier2020, Pillitteri2022HD189, Tristan2023, Ilin2022, Feinstein2022}. 

For 10 observations, we generated two identical sets of each light curve: one that could have flares and one without a chance of flares. Then, for each point in the flaring light curve, before averaging observations, we generated a random number between 0 and 1 from a uniform distribution. If that number was less than or equal to the 
less than or equal to the probability that an X-ray flare will occur within the span of the time bin, that point was marked as the start of a flare. We then drew from a distribution of flare durations for each star, and masked 
the light curve for that duration, rounded up to the nearest 500 second bin. 
For HD~189733, we used a flare rate of 0.065 per ks \citep{Pillitteri2022HD189}. For AU~Mic, we divided the number of X-ray flares detected by \citet{Tristan2023} by the total time they spent observing in X-ray to obtain a value of 0.1295 flares per ks. The flare durations for each star were chosen randomly from the durations present in these respective studies. 

We then averaged the 10 observations and completed the analysis as in our previous experiments, for both simulated datasets. With 100 repeated experiments we found that, for HD~189733~b, the median p-value for simulated datasets with and without flares both indicated a $\geq$4.4$\sigma$ detection. For AU~Mic~b, the median p-values differed by less than 0.5$\sigma$, with the flaring dataset detectable at a $1.8\sigma$ significance. This indicates that, while flares have a large effect on single observations, they are not likely to prevent a transit from being detectable with 10 averaged observations.

\section{Discussion}
\label{sec:disc}

Figure~\ref{fig:TranDepScatter} (left) shows the optical transit depth versus apparent X-ray flux for all the transiting planets in our sample, and Figure~\ref{fig:TranDepScatter} (right) shows the planetary optical radius versus orbital period. The top 15 transit candidates are highlighted in both figures. All of the top 15 planets have orbital periods less than 10 days 
and radii between 4 and 23 times that of earth, indicating that next generation of X-ray observatories will be able to study both Neptunes and hot Jupiters. The ages of the top 15 host stars, where specified in the NEA, range from 0.017 to 7.2 Gyr. About half of them have system ages of less than 1 Gyr, suggesting that the host stars could be within $\sim$1 magnitude of their peak X-ray brightness. The proximity of these planets to their host stars as well as their large size, implying their gaseous nature, suggests that they are likely to have significant atmospheric photoevaporation and extended apparent radii in X-ray observations.

The planets that are most likely to be observable, even if they do not have extended atmospheres, are gas giants with short orbital periods: the hot Jupiters 
HD 189733 b, HIP 65 A b, and WASP-93 b with orbital periods of 2.2, 0.98, and 2.73 days respectively \citep{Stassun2017, Nielsen2020, Hay2016}. HD 189733 and HIP 65 A are both known to have an M-dwarf companion: HD 189733B with a separation of 216 au ($\sim$11", \citet{Bakos2006}), and HIP 65 B with a separation of 245 au (3.95", \citet{Nielsen2020}). HIP 65 B will be resolvable from its companion star with the proposed angular resolution of AXIS ($\lesssim 2''$), but not \textit{NewAthena} ($\lesssim 10''$). Thus transits of HIP 65 A b, especially those observed with \textit{NewAthena}, may be more difficult to detect than our simulations indicate and will require adjustments for dilution. The fourth candidate target, AU Mic b, is a Neptune-like planet orbiting one of the youngest and most nearby M-dwarf stars, with a short orbital period of 8.5 days \citep{Cale2021}. In this study, we used an X-ray flux of 4.7$\times10^{-11}$ erg\,cm$^{-2}$\,s$^{-1}$ measured with ROSAT \citep{Hunsch1999}. More recent XMM-Newton studies of AU Mic have suggested that the quiescent X-ray flux could be lower by about a factor of two \citep[e.g.][]{Spinelli2023, Poppenhaeger2022, Tristan2023}. As such, our results represent an optimistic estimate of the AU Mic count rate in AXIS and \textit{NewAthena} and real observations may find the transit more difficult to detect than our results indicate, depending on the level of activity of the star at the time of observation.

\subsection{Limitations of this Study}
\label{ssec:Limitations}

The largest source of uncertainty in this study is the order-of-magnitude dispersion in $L_{\rm X}/L_{\rm bol}$ versus age and rotation relations observed by W11, J12, W18, and similar studies \citep[e.g.][]{Pizzolato2003,Stelzer2016,Johnstone2021,Magaudda2022,R-Y2023}. Only one of the top five planetary systems in our study had an X-ray luminosity that was estimated from these relationships. That was our third highest ranked transit target, WASP-93~b, for which we estimated a stellar X-ray flux of $4\times10^{-13}$ erg\,cm$^{-2}$\,s$^{-1}$ using the W18 relation with a rotation period of 1.45$\pm$0.45 days \citep{Hay2016}. This target may move up or down on the ranked lists presented in this work, pending future observations of its X-ray luminosity. The same is true for four other targets in the top 15 observational candidates with X-ray luminosities estimated via J12 or W18.

Some uncertainties also arise from stellar age measurements. The relationship from J12 is derived from a large sample of stars, but whose ages are all below 1 Gyr. This could lead to error when the same relationships are applied to older stars. For example, \citet{Booth2017} found a steeper decrease in X-ray emission with age, with a smaller sample of stars older than 1~Gyr. In addition, measurements of field star ages often have large uncertainties, introducing further error into the X-ray luminosities we estimated using the J12 relationship. None of the 44 systems with X-ray luminosity estimated from a randomized stellar age yielded detectable transits, so there are no false-positives coming from this population. However, there is still the possibility of false-negatives, as our choices could  
significantly underestimate the X-ray flux of some of these stars. We consider it unlikely that field stars, which have had sufficient time to wander away from their cluster of birth, are in the saturated phase of $L_X/L_{\rm bol} \sim 10^{-3}$. Choosing a maximum value of $L_X/L_{\rm bol} = 10^{-5}$ we find that, of those field stars that are nearby enough to achieve a similar brightness to some of the dimmer targets in our list of the top 15 transit targets, all have planets that are smaller $(R_p/R_* < 0.1)$, making them unlikely to yield high signal transit depths. While not impossible, we consider it unlikely that a stand-out transit candidate has been missed.

For most of the stars with a measured X-ray flux, very little is known about the magnitude of long timescale variations. As an example, the X-ray luminosity of the Sun varies by over an order of magnitude over the course of its 11-year activity cycle \citep{Peres2000}. Additionally, X-ray flux measurements in the literature capture individual snapshots, where a star may be in a moment of prolonged outburst \citep[e.g.][]{Salz2019}. This vast uncertainty inspires us to report more than the best few candidates for X-ray transit measurement, as they may be good targets for observation depending on the level of stellar activity.
Further X-ray studies of planet-hosting stars are critical to better characterizing $L_{\rm X}/L_{\rm bol}$ relationships and measuring X-ray transits. 

Short flares and active regions can also lead to extreme complications in the shape of the X-ray transit. Contamination of the transit light curve can be far more pronounced in the X-ray than the optical, leading to significantly deeper transits if the planet happens to occult an active region of the star \citep{Llama2015}. On the other hand, if a short flare occurs or an active region rotates into view over the course of the transit, the transit depth will appear shallower. Averaging the light curves of multiple transit observations mitigates correlated noise and deviations of transit shape, as a star is unlikely to vary in the same way during different observations. \citet{Llama2016} demonstrated that, in the similarly extreme regime of Ly-$\alpha$ transit measurements, the correct transit depth can be recovered 90\% of the time by averaging over 10 transits. Since the majority of our experiments utilized 10 averaged transit observations, the effects of correlated noise and star spots should not significantly change our principal identification of top transit targets. Our results in section \ref{ssec:Flares} are in agreement, as the detection significance of HD 189733 b and AU Mic b are similar with and without simulated flares during the observations.

In analysis of real data, the methods will need to be optimized for the specific properties of the equipment used for the measurement.  
In this work, we simulated transit light curves by sampling from a Poissionian distribution, assuming that the data points are uncorrelated. In reality, data points may contain systematic correlations arising from the instrument. An additional source of bias could arise from observational sampling limitations. As an example, observations from telescopes in low-Earth orbit (e.g., Swift,  XRISM, and AXIS [proposed]) are frequently executed in short exposures $\sim 30-90$ minutes depending on the target coordinates. This sampling timescale is similar to the timescale of typical short-period planet transits ($\sim 3$ hours), introducing a source of systematic bias. 
The resulting gaps in the transit light curve, in addition to gaps arising from flare removal, can leave a substantial fraction of the transit duration missing. Gaps also affect the pre- and post-transit light curves, which provide an important baseline for studying underlying stellar variations and determining the transit depth.
The simulations also assumed that any background signal would be negligible. 
All of these factors need to be considered by individual telescope teams and observers. 

It is an active area of investigation to understand to what extent tidal and magnetic interactions within systems containing hot Jupiters can lead to increases in stellar rotation periods to the point that empirical relationships for X-ray activity may no longer apply \citep{Cuntz2000, Scharf2010, Poppenhaeger2014}. The effect of these interactions is to generally increase the X-ray luminosity of the host star relative to what might be expected. This may boost the likelihood of transit detection for some of the transit targets in our study for which the intrinsic X-ray luminosity of the host star was inferred from other stellar parameters. Considering that this is still an open question, we did not include star-planet interactions when estimating stellar X-ray luminosities.

\subsection{Comparison to Other Works}
\citet{Foster2022eROSITA} present eROSITA X-ray studies of planet hosting stars, some of which we included in our study. They use the measured X-ray luminosities to predict mass-loss rates, identifying a few transiting exoplanets with high mass-loss rates which would be interesting for further study: TOI-251~b, GJ~143~b, K2-198~bcd, and K2-240~bc. TOI-251~b and GJ~143~b were both in our sample. However, simulated transit light curves with 10 averaged observations and a radius 150\% larger than optical showed that their transits were too shallow to appear through statistical noise, even with a heavily extended apparent X-ray radius. The K2-198 system was not in our sample due to our cut of dim systems farther than 100 pc. Although we did not simulate their transits, K2-198 b, c, and d will likely not have detectable transits in AXIS or \textit{NewAthena} as the largest planet of the three is similar to AU Mic b in radius, but its star is significantly dimmer than AU Mic \citep{Foster2022eROSITA}. Similarly, GJ~143~b and c both have a much smaller $\frac{R_p}{R_*}$ than any of our potentially observable targets \citep[$\sim0.03$ for both planets,][]{DiezAlonso2018}, and the stellar X-ray flux of $\sim6\times10^{-14}$ erg\,cm$^{-2}$\,s$^{-1}$\citep{Foster2022eROSITA} is not large enough for such small transits to be detectable. 

\citet{Foster2022} also use eROSITA data to estimate mass-loss rates for exoplanets. They note DS~Tuc~A~b and WASP-180~A~b, two high mass-loss transiting systems in binaries, as interesting systems for follow-up observations with  \textit{NewAthena}. Our study shows that both of these planets may be detectable in \textit{NewAthena}, provided at least 10 observations and an apparent X-ray radius at least 50\% larger than optical. However, as both planets have known X-ray bright companions with separations less than 6" \citep{Temple2019, Foster2022}, the transits may be more shallow and difficult to detect than our simulations indicate. 
\citet{Foster2022} additionally simulate stellar spectra during a transit of a hypothetical hot Jupiter through the \textit{NewAthena} X-IFU instrument, suggesting that absorption signatures in outer exoplanet atmospheres may be detectable. Combining the X-IFU data with the power of the WFI to detect transits at a high significance will thus allow \textit{NewAthena} to study exoplanet atmospheres from a new perspective in the electromagnetic spectrum.

\section{Conclusions}
\label{sec:conc}

In this study we have estimated the feasibility for future generations of X-ray telescopes with large collecting areas, such as AXIS and \textit{NewAthena}, to detect exoplanet transits. We collected a large sample of nearby transiting exoplanets with data from the NASA Exoplanet Archive. For those missing X-ray observations, we estimated the X-ray luminosity of each host star using empirical relationships between X-ray luminosity and stellar rotation, age, and color. We modeled the transits as they would appear in AXIS and \textit{NewAthena} observations and used the simulated light curves to determine the probability for each transit to be successfully detected. We then identified the top 15 transit targets, and ordered them according to their probability of detection. In all cases, we found that stacking multiple transit light curves is required to obtain transit detections to $\geq3$-sigma level significance.

Our primary results indicated that, in the absence of atmospheric escape, a small number of exoplanet transits could be detectable by AXIS and \textit{NewAthena}: HD~189733~b, HIP~65~A~b, WASP-93~b, and possibly AU~Mic~b. WASP-93~b is the only candidate of these four for which we did not find prior X-ray measurements, so the X-ray luminosity was inferred from its rotation period \citep{Hay2016} using the relationship from \citet{Wright2018}. The order of magnitude uncertainties on these X-ray luminosity relations can easily elevate or remove this planetary transit target from future consideration.

It is highly probable that, due to atmospheric escape, the apparent radius of a gaseous short-period planet will be significantly larger in the X-ray than it is in the optical. This has been observed to be the case for HD~189733~b \citep{Poppenhaeger2013}, which has a relatively small mass-loss rate. All of our top 15 transit targets orbit their host stars within 10 days and are large enough to have atmospheres. Thus many of them could be experiencing atmospheric escape that might be detectable in the X-ray.
We repeated our experiments to discern the detection probability of detecting transits under the assumption of larger effective X-ray radii. 
With an X-ray radius that is 2.5 times larger than the optical, all but one of the top 15 candidates would be detectable at a $>3\sigma$ significance by \textit{NewAthena}.

Stellar variation, and the order of magnitude dispersion observed in demographics studies of stellar X-ray luminosities, is the primary limitation for this study. 
This can be addressed by X-ray study and monitoring of the planet-hosting stars in our sample. Additionally, the effective area both AXIS and the \textit{NewAthena} WFI will enable more extensive and precise demographic study of stellar activity. The high resolution imaging capabilities of AXIS relative to \textit{NewAthena} will also make it particularly more effective for constraining the activity-age relations for young star systems, including M dwarfs \citep{Corrales2024}.

Our work demonstrates that the opportunity for studying exoplanet atmospheres from the X-ray perspective  will increase substantially with missions like AXIS and \textit{NewAthena}. This unprecedented X-ray capability will enable us to better characterize exoplanet atmospheres in their current and prior states, assisting research in various fields of exoplanet study from locating habitable worlds to tracing the formation of planetary systems.

\begin{acknowledgements}

This research has made use of data obtained from the 4XMM XMM-Newton Serendipitous Source Catalog compiled by the 10 institutes of the XMM-Newton Survey Science Centre selected by ESA. 
This research has made use of the NASA Exoplanet Archive, which is operated by the California Institute of Technology, under contract with the National Aeronautics and Space Administration under the Exoplanet Exploration Program. 
This paper makes use of data from the first public release of the WASP data \citep{Butters2010} as provided by the WASP consortium and services at the NASA Exoplanet Archive, which is operated by the California Institute of Technology, under contract with the National Aeronautics and Space Administration under the Exoplanet Exploration Program. 
These data are made available to the community through the Exoplanet Archive on behalf of the KELT project team. 
This research has made use of the SIMBAD database, operated at CDS, Strasbourg, France. 
This work has made use of data from the European Space Agency (ESA) mission {\it Gaia} (\url{https://www.cosmos.esa.int/gaia}), processed by the {\it Gaia} Data Processing and Analysis Consortium (DPAC, \url{https://www.cosmos.esa.int/web/gaia/dpac/consortium}). Funding for the DPAC has been provided by national institutions, in particular the institutions participating in the {\it Gaia} Multilateral Agreement. 
    
\end{acknowledgements}

\facility {Exoplanet Archive, XMM-Newton, Chandra, \textit{Gaia}, ROSAT, eROSITA} 

\software {Astropy \citep{Astropy2013, Astropy2018, Astropy2022}, XSPEC \citep{Arnaud1996}, batman \citep{Kreidberg2015}, Science Analysis System, scipy \citep{Scipy}, numpy \citep{numpy}}

\bibliography{main}{} 
\label{sec:bib}
\bibliographystyle{aasjournal}

\end{document}